\documentclass[aps,prl,twocolumn,altaffilletter,nolongbibliography,numerical,flushbottom,secnumarabic,superscriptaddress,floatfix,10pt]{revtex4-2}

\usepackage{graphicx}
\usepackage{braket}
\usepackage{booktabs}  
\usepackage{comment}
\usepackage[normalem]{ulem}
\usepackage[title]{appendix}
\usepackage{amsmath,mathtools,amsthm,amssymb,pifont}
\usepackage[utf8]{inputenc}
\usepackage[american]{babel}
\usepackage{graphicx,xcolor,bbold,titlesec}
\usepackage{MnSymbol}

\usepackage{float}

\usepackage{mathtools}

\usepackage{tikz}
\usetikzlibrary{decorations.pathreplacing, calc}

\usepackage[breaklinks, pdftex, hyperfootnotes=true, pdfpagelabels, bookmarks, pageanchor]{hyperref}
\hypersetup{%
	colorlinks=true, linktocpage=true, pdfstartpage=1, pdfstartview=FitH, pdfborder={0 0 0},%
	breaklinks=true, pdfpagemode=UseNone, pageanchor=true, pdfpagemode=UseOutlines,%
	plainpages=false, bookmarksnumbered, bookmarksopen=true, bookmarksopenlevel=1,%
	hypertexnames=true, pdfhighlight=/O,
	urlcolor=red, linkcolor=red, citecolor=red,
	}
\usepackage{orcidlink}
\usepackage{tikz,ifthen}
\usepackage{bbold}
\usepackage{tikz-network}
\usetikzlibrary{patterns,decorations.pathreplacing,calligraphy}
\usetikzlibrary{shapes,arrows.meta,decorations.pathmorphing}

\graphicspath{{pictures}}

\newcommand{\Tr}{{\text{Tr}}}

\newcommand{\Wg}{{\text{Wg}}} 

\newcommand{\kket}[1]{\left. \left| #1 \right\rangle \right\rangle}
\newcommand{\bbra}[1]{\left\langle \left\langle #1 \right|\right.}
\newcommand{\bbrakket}[2]{\left\langle \left\langle #1 | #2 \right\rangle \right\rangle}

\begin{document}

\newcommand{\titleinfo}{Correlations, Spectra and Entanglement Transitions \\ in  
Ensembles of Matrix Product States }
\title{\titleinfo}

\author{Hugo Lóio~\orcidlink{0000-0002-4011-8180}}
\affiliation{Laboratoire de Physique Th\'eorique et Mod\'elisation, CNRS UMR 8089,
CY Cergy Paris Universit\'e, 95302 Cergy-Pontoise Cedex, France}

\author{Guillaume Cecile~\orcidlink{0000-0002-2076-6236}}
\affiliation{Institute for Quantum Physics, University of Hamburg, Luruper Chaussee 149, 22761 Hamburg, Germany}
\affiliation{Laboratoire de Physique Th\'eorique et Mod\'elisation, CNRS UMR 8089,
CY Cergy Paris Universit\'e, 95302 Cergy-Pontoise Cedex, France}

\author{
Sarang Gopalakrishnan
}

\affiliation{Department of Electrical and Computer Engineering, Princeton University, Princeton NJ 08544, USA}

\author{Guglielmo Lami~\orcidlink{0000-0002-1778-7263}}
\affiliation{Laboratoire de Physique Th\'eorique et Mod\'elisation, CNRS UMR 8089,
CY Cergy Paris Universit\'e, 95302 Cergy-Pontoise Cedex, France}

\author{Jacopo De Nardis~\orcidlink{0000-0001-7877-0329}}
\affiliation{Laboratoire de Physique Th\'eorique et Mod\'elisation, CNRS UMR 8089,
CY Cergy Paris Universit\'e, 95302 Cergy-Pontoise Cedex, France}

\begin{abstract}
We investigate ensembles of Matrix Product States (MPSs) generated by quantum circuit evolution followed by projection onto MPSs with a fixed bond dimension $\chi$. Specifically, we consider ensembles produced by: (i) random sequential unitary circuits, (ii) random brickwork unitary circuits, and (iii) circuits involving both unitaries and projective measurements. In all cases, we analyze the spectra of the MPS transfer matrices and relate them to the spreading of mutual information in the MPS state. We demonstrate how different features of the spectral density correspond to distinct types of circuits, revealing that these MPS ensembles retain crucial physical information about the underlying microscopic dynamics. Notably, in the presence of quantum monitoring, we show the existence of a measurement-induced entanglement transition (MIPT) in MPS ensembles, with the averaged dimension of the transfer matrix’s null space serving as the effective order parameter.
\end{abstract}

\maketitle

\emph{Introduction. ---} 
Simulating highly entangled quantum states is one of the central challenges in condensed matter physics, quantum information science, and statistical physics.
In one-dimensional systems with bounded entanglement~\cite{MatrixProductStateRepresentations,Verstraete2008,PhysRevB.73.094423,Schollwck2011}, tensor networks provide a powerful and efficient framework for representing quantum many-body states classically, and also for preparing them efficiently
on digital quantum platforms~\cite{piroli2021quantum,malz2024preparation,PRXQuantum.5.030344,david2024preparing,zhang2024yifan}.
In such cases, the bond dimension $\chi$, which defines the size of each tensor in the network, can be kept relatively small and independent of the system size. Ground states of local Hamiltonians are typically weakly entangled, but in more general settings (e.g.\ when describing nonequilibrium dynamics~\cite{Nahum2017,Aniket2023,Calabrese2016,Fagotti2008}) the bond dimension $\chi$ must scale exponentially with the system size.
This exponential growth makes exact classical tensor network representations infeasible. 
Approximate representations, in which a state is projected onto a manifold of tensor network states with a fixed $\chi$, have been explored in the literature~\cite{leviatan2017quantum,Cecile2024,Begui2024,PhysRevB.97.024307,PRXQuantum.4.020304}, but the extent to which such approaches can fully capture the underlying quantum dynamical process remains unclear.

\begin{figure}[t!]
  \centering
  \includegraphics[width=0.55\columnwidth]{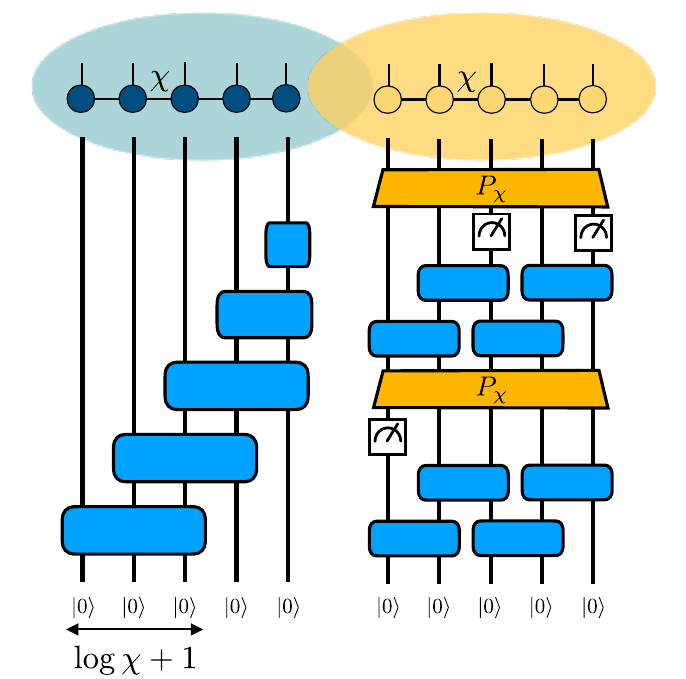}
  \caption{
 Illustrative overview of this work. We study ensembles of Matrix Product States (MPSs) characterized by a fixed bond dimension $\chi$, including: (left) random MPSs generated by a staircase quantum circuit; and (right) MPSs produced by quantum circuits composed of random two-qubit gates and, optionally, single-site measurements. In the latter case, the wave function is projected onto the MPS manifold with bond dimension $\chi$ and renormalized after each layer (orange shapes). While exhibiting distinct features, such as specific spectral and correlation structures at finite $\chi$, we show that the two ensembles share several common properties in the large-$\chi$ limit (with $\chi/2^N \to 0$).  
  }
  \label{fig_work_scheme}
\end{figure}
In this work, we introduce ensembles of matrix product states (MPSs) that, remarkably, despite having a finite bond dimension, can mirror most of the features of quantum dynamics and, in particular, of entanglement spreading and entanglement transitions. 
We consider ensembles generated by a paradigmatic family of chaotic many-body systems: namely, random quantum circuits~\cite{Fisher2023,2023Google}, where each two-site gate is randomly selected from the Haar unitary ensemble, see Fig. \ref{fig_work_scheme} and where the MPS is projected back on the space of finite bond dimension $\chi$ on each layer. {We consider both unitary circuits and hybrid circuits in which projective measurements are introduced at a fixed rate \(p\)~\cite{li2019measurementdrivenentanglement,chan2018unitary,PhysRevX.9.031009,vijay2012stabilizing,katz2006coherent,campagneibarcq2016observing,bauer2015computing,PhysRevB.106.L220304}.  Our main result is that, even if one truncates the dynamics to a space of fixed $\chi$, the MPS ensembles retain information about the structure of the underlying dynamics. For unitary dynamics, we show that random sequential circuits and random brickwork circuits yield transfer matrices featuring a spectral gap that approaches a universal, $\chi$-independent value at large $\chi$, but with different spectral densities (defined as the eigenvalue probability density in the complex plane).  It is generally believed that spatial correlations in MPSs are encoded in the transfer matrix spectrum, where the largest eigenvalue is associated with the correlation length {$\xi$} \cite{PhysRevX.8.041033,PRXQuantum.4.030330,PhysRevB.92.235150,zauner2015transfer,PhysRevLett.132.086503,PhysRevLett.123.250604}. {Naively, this would suggest that sequential and brickwork circuits, once projected on MPS manifolds, generate the same correlations}. We demonstrate that this interpretation is incomplete, as the full spectral density of eigenvalues participates in the spatial correlations. We find that correlations are finite over a region of size $\xi_{\mathrm{eff}} \sim \log \chi^\alpha$ where $\alpha$ is distinct for different circuits, reflecting the fact that these circuits have distinct entanglement dynamics.} Finally, we show that in the presence of projective measurements, the spectrum of the transfer matrix is itself changed, with the appearance of midgap states for any value of measurement rate, and, mostly importantly, with a finite null space for any value of $\chi$, whenever the measurement rate exceeds the critical measurement-induced phase transition (MIPT) value. The latter is responsible for the emergence of the area-law phase, where $\xi_{\mathrm{eff}} \sim \log \chi^0$.


Our findings elucidate how long-range correlations can also arise from transfer matrices with finite spectral gaps and demonstrate how key properties of truly quantum many-body systems are mirrored in the structure of MPS ensembles with modest values of bond dimensions.

\emph{Preliminaries: MPSs, transfer matrices and correlation lengths ---} 
MPSs
are characterized by $d$ matrices of size $\chi \times \chi$ for each qudit, with local dimension $d$. A generic MPS with bond dimension $\chi$ can therefore be written as 
\begin{equation}
\begin{split}
    | \psi \rangle =  
    \begin{gathered}
        \includegraphics[width=0.45\columnwidth]{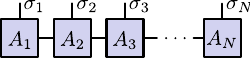}
    \end{gathered}
\end{split}
    \,.
\end{equation}
MPSs can be efficiently contracted by choosing an appropriate gauge, characterized by a center of orthogonality. In this gauge, all tensors are either left- or right-isometries, except for the tensor at the orthogonal center~\cite{Schollwock_2011}. Expectation values of correlation functions can be expressed in terms of the transfer matrix:
\begin{equation}\label{eq_transfer_matrix}
    \mathcal{T}_i = \sum_{\sigma=1}^d (A_i^{\sigma})^* \otimes A_i^{\sigma} 
    = 
    \begin{gathered}
        \includegraphics[width=0.12\columnwidth]{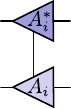}
    \end{gathered}
    \, .
\end{equation}
In a translationally invariant MPS, $A_i = A$ and $\mathcal{T}_i = \mathcal{T}$ for all $i$. 
For instance, the connected correlation function of a local operator $O_i$ between positions $i$ and $i+r+1$ can be expressed in terms of the left and right eigenvectors, $\bbra{l_m}$ and $\kket{r_m}$, and the corresponding eigenvalues $\lambda_m$ of the transfer matrix $\mathcal{T}$ as~\cite{Schollwock_2011}:
\begin{equation}\label{eq:correlation1}
    \langle O_i O_{i+r+1} \rangle^c = \sum_{m > 1} \lambda_m^r \bbra{l_1} \mathcal{T}^O \kket{r_m} \bbra{l_m}  \mathcal{T}^O \kket{ r_1 },
\end{equation}
where, similarly to Eq.~\eqref{eq_transfer_matrix}, we define $\mathcal{T}^O = \sum_{\sigma} O^\sigma (A^{\sigma})^* \otimes A^{\sigma}$. Normalization imposes $\lambda_{1} = 1$, while all other eigenvalues lie within the unit disk on the complex plane: $|\lambda_m| < 1$, for $m>1$.
This typically motivates the use of a single leading eigenvalue $\lambda_{\rm max} = \underset{\lambda \in \{\lambda_m , \, m>1\}}{\mathrm{arg\,max}} \,|\lambda| $ to define the correlation length $\xi$, thereby reducing the sum in Eq.~\eqref{eq:correlation1} to the leading term~\cite{Schollwock_2011}:
\[
\langle O_i O_{i+r+1} \rangle^c \sim e^{-r / \xi}, \quad \text{with } \xi = -  1/{\log |\lambda_{\rm max}|}.
\]
The spectral gap, defined as $1 - |\lambda_{\rm max}|$, characterizes the criticality of the state. For gapless ground states, it is known that the gap closes in the limit of large $\chi$, such as the correlation length diverges as  $\xi \sim \chi^{\kappa}$, with $\kappa$ an exponent directly related to the underlying field theory \cite{PhysRevLett.102.255701,PhysRevB.78.024410,PhysRevLett.129.200601,PhysRevLett.123.250604,2411.03954}. 
The phenomenology is instead much different in the case of MPSs approximating volume-law states, as we shall show in the following. 

\begin{figure}[t]
  \centering
  \includegraphics[trim=5 5 5 5, clip, width=\columnwidth]{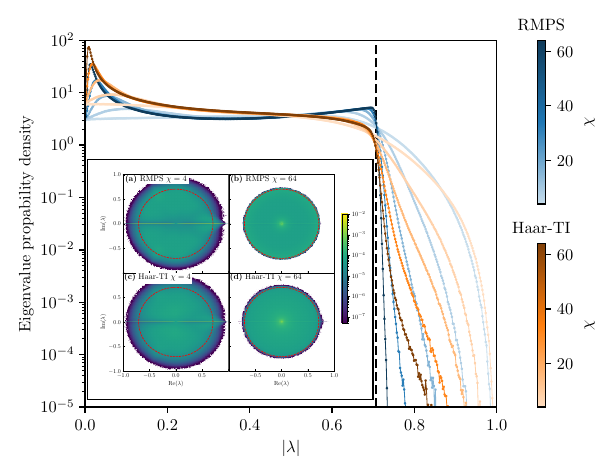}
  \caption{
    Radial spectral density (eigenvalues probability density) of the transfer matrix at fixed bond dimension $\chi$ obtained from an RMPS or a translational-invariant Haar random circuit.  
    The radial densities are obtained by integration over the angles of the probability density of eigenvalues in the complex plane, and it shows the convergence to two different asymptotic densities, which share the same support of size $1/\sqrt{d}$  (indicated for $d=2$ by the dashed vertical line) (eigenvalues at $|\lambda| = 1$ have been removed). 
  \textit{Inset}: the full spectral density in the complex plane is shown, for specific values of $\chi$ (top is for RMPSs and below for a translational-invariant Haar random circuit). 
  }
  \label{fig_RMPS_TI_Haar_eigs}
\end{figure}

\begin{figure*}
    \centering
    \includegraphics[width=\linewidth]{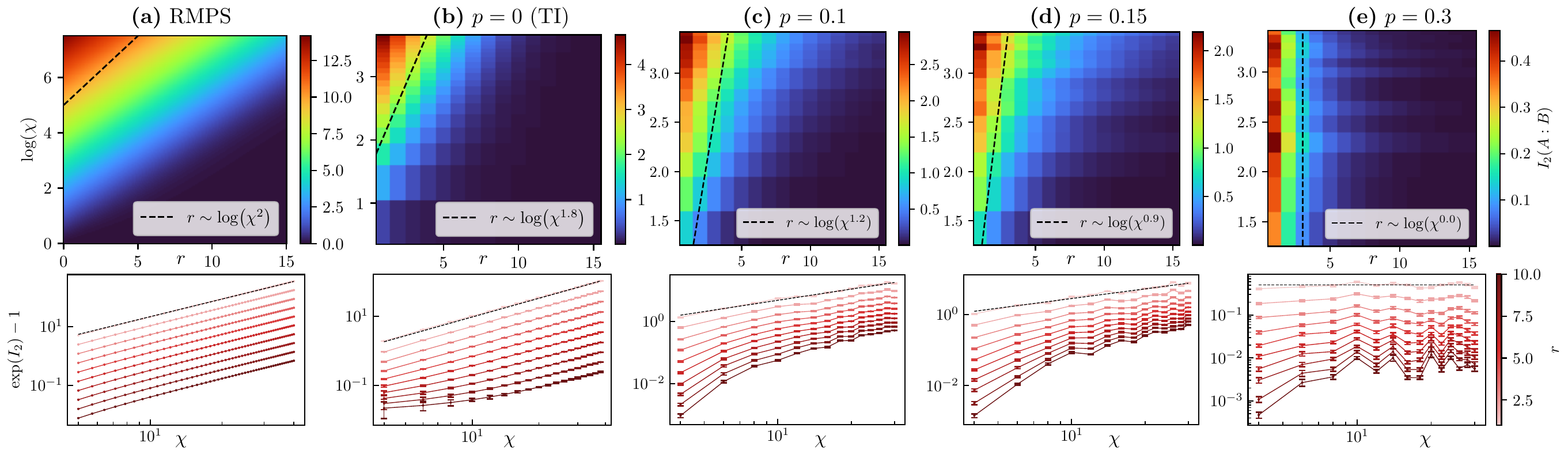}
    \caption{
    Top row: density plots for the Rényi-2 mutual information as a function of $r$ and $\log(\chi)$. The dashed line is given by the effective correlation length $\xi_{\rm eff}
\sim \log(\chi^\alpha)$, where the exponent is taken from fitting the slope of the plots in the bottom row, for $r = 1$, also shown in dashed.
    Bottom row: exponential of the mutual information as a function of $\chi$, for fixed $r \in 1-10$. The data in (a) and (b) are for an infinite circuit, and data in (c) (d) and (e) are for a monitored circuit of size $N=60$. 
    }
    \label{fig_mutual_info}
 \end{figure*}

A simple class of MPSs that flow to volume-law states are random MPSs (RMPSs) \cite{PhysRevA.81.032336,PhysRevA.82.052312,Lancien2022,PRXQuantum.2.040308,PhysRevResearch.3.L022015,PRXQuantum.4.030330,lami2024quantumstatedesignsclifford,lami2024anticoncentrationstatedesignrandom} which can be obtained by sequential unitary circuits, see Fig. \ref{fig_work_scheme}, where each gate is a Haar random gate acting on $\log_d \chi + 1$ sites.  As RMPSs have entanglement entropy of order $\log \chi$, they flow to high-entanglement states in the limit $\chi \to \infty$. However, their spectral density converges to a well-defined limiting shape in this limit. Since the transfer matrix \eqref{eq_transfer_matrix} in this case can be interpreted as a \textit{random quantum channel} \cite{Wolf2008,kukulski2021generating,gonzalez2018spectral,PhysRevLett.123.140403,MatsoukasRoubeas2024quantumchaos,PhysRevB.102.134310} given by the sum of $d$ random Kraus operators acting on a Hilbert space of dimension $\chi^2$, it can be argued that its spectrum in the limit of a large dimension $\chi$ is confined within a circle of radius $1 / \sqrt{d}$ \cite{kukulski2021generating}, which we indeed confirm in Fig. \ref{fig_mutual_info} and \ref{fig_RMPS_TI_Haar_eigs}. 
The existence of a finite, maximal correlation length in the MPS in the limit $\chi \to \infty$ naively appears to conflict with the growth of the bipartite entanglement in this limit, as $S = \log \chi$. 
This apparent contradiction is resolved by reconsidering the largest eigenvalue approximation.
In the following section, we show that while the spectral gap is a reliable measure of long-range correlations in critical or area-law states, this does not apply to MPSs that flow to volume-law states at large $\chi$.
In this case, the entire sum over the eigenstates of the transfer matrix contributes to the spatial correlations.

\emph{Effective correlation length in translational invariant MPSs and in RMPSs ---} We employ the Rényi-$k$ mutual information to quantify the correlations between two semi-infinite subsystems, denoted as $A$ and $B$ separated by distance $r$, in the 1D geometry (the extension to the 2D case can be done analogously). 
The Rényi-$k$ mutual information between the two blocks is defined as
$ I_k(A:B) = S_k(\rho_A) + S_k(\rho_B) - S_k(\rho_{A \cup B})$, where $S_k$ corresponds to the Rényi entropy of order $k$ (the Von Neumann entropy is found in the limit $k \rightarrow 1$.) 
For $k \geqslant 2$, the mutual information can be expressed as 
$I_k(A:B) = \frac{1}{k-1} \log \left[\frac{\Tr(\rho_{A \cup B}^k)}{\Tr(\rho_A^k) \Tr(\rho_B^k)} \right]$.  For translation invariant MPSs, i.e. $A_i = A$, the expression for the $I_k$ can be further simplified.
Computing the terms $\Tr(\rho^k_X)$ involves taking a product of two distinct transfer matrices in $k$ replicated bond space, one in the region $X$ and the other in $\overline{X}$.
The transfer matrices are explicitly expressed as 
\begin{equation}
\resizebox{!}{0.135\columnwidth}{
    $
    \displaystyle
    \mathcal{T}_\alpha^{(k)} =
    \sum_{\boldsymbol{\sigma} \boldsymbol{\sigma'}}
    \left(\prod_{i=1}^k\delta_{\sigma_i \sigma'_{\alpha(i)}}\right)
    \left(\bigotimes_{i=1}^k A^{\sigma_i} \otimes
     \left(A^{\sigma'_i}\right)^*
     \right)
     =
    \begin{gathered}
        \includegraphics[width=0.15\columnwidth]{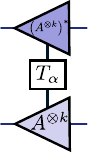}
    \end{gathered}
    \, ,
    $
    }
\end{equation}
where $\sum_{\boldsymbol{\sigma} \boldsymbol{\sigma'}} = \sum_{\sigma_1 \sigma_2 \dots \sigma_k, \sigma'_1 \sigma'_2 \dots \sigma'_k}$ and $\alpha$ is a permutation with corresponding matrix representation $T_\alpha$.
Note that $\mathcal{T}_\alpha^{(k)}$ corresponds to a $k$-replicated version of Eq.~\eqref{eq_transfer_matrix}, with an additional permutation over the physical indices, and therefore has dimension $\chi^{2k}$.
There is a permutation freedom in the ordering of the replicated MPS tensors, which we fix by setting $\alpha = e$, the identity permutation, for the transfer matrix $\mathcal{T}_{(1)} \equiv \mathcal{T}^{(k)}_e $ of the region $X$.
Consequently, in region $\overline{X}$, the correct permutation is $\alpha = C_k =  (1 2 \dots k)$, i.e. a $k$-cycle, yielding the transfer matrix $\mathcal{T}_{(2)} \equiv \mathcal{T}^{(k)}_{C_k}$.
The transfer matrix of region $\overline{X}$ has eigen decomposition over a set of $\chi^{2k}$ left and right eigenvectors, 
$\mathcal{T}_{(2)} = \sum_{m=1}^{\chi^{2k}} \Lambda_m \kket{R_m} \bbra{L_m}$. 
The set of eigenvalues $\{\Lambda_m$\} can be fully computed as products of $k$ eigenvalues of $\{\lambda_m\}$, therefore the conditions $\Lambda_1 = 1$ and $|\Lambda_{m > 1}| < 1$ still hold.
As for the transfer matrix $\mathcal{T}_{e}^{(k)}$ in the region $X$, we will only be concerned with its leading left and right eigenvectors, which we denote by $\bbra{L}$ and $\kket{R}$ respectively, corresponding to eigenvalue 1.

The mutual information can then be expressed as
\begin{equation}
\label{eq_mutual_info_TI}
    I_k(A:B) =
    \frac{1}{k-1}
    \log
    \left[ 
    1 + \sum_{m>1}^{\chi^{2k}} \Lambda_m^r
    \frac
    { \bbrakket{L}{R_m} \bbrakket{L_m}{R} }
    { \bbrakket{L}{R_1} \bbrakket{L_1}{R} }
    \right],
\end{equation}
where the boundary vectors of $\mathcal{T}_{(1)}$, $ \bbra{L}$ and  $\kket{R}$, and of $\mathcal{T}_{(2)}$ , $\bbra{L_1}$ and $\kket{R_1}$, are the ones for which the $k$-replica normalization conditions $\Tr[ \rho^k] = \Tr[\rho]^k = 1$ hold (the full state $\rho$ is pure). As the eigenvalues are exponentiated by $r$, analogously to the 2-point correlation case, at large $r \gg 1$ the leading eigenvalue $\Lambda_{\rm max} = \underset{\Lambda \in \{\Lambda_i, i>1\}}{\mathrm{arg\,max}} \,|\Lambda| $ will dominate the sum and set the correlation length which is defined by $\xi = - 1 / \log|\Lambda_{\rm max}|$.
However, at finite $r$, the full sum in eq. \eqref{eq_mutual_info_TI} cannot be neglected: if the sum over the $\chi^{2k} - 1$ converges at large $\chi$ clearly only the leading eigenvalue dominates, but if not, we shall expect that this sum at large $\chi$ is given by $\sim e^{- r /\xi} \times {\rm poly} (\chi)$. In this case, the effective correlation length is instead given by $\log {\rm poly} (\chi) $. 

The latter phenomenon can be shown directly in the RMPS ensemble. 
Although these are not translational invariant, we make the assumption $\mathbb{E}\log[\Tr(\rho_X^k)] \approx \log[\mathbb{E}\Tr(\rho_X^k)]$, stating that the statistical fluctuations of $\Tr(\rho_X^k)$ are small, where $\mathbb{E}$ refers to the Haar average of the unitary matrices that form the MPS ensemble.
After averaging, the replicated transfer matrices become translational invariant and Eq.~\eqref{eq_mutual_info_TI} still applies.
The averaging can be performed analytically with Weingarten calculus \cite{collins2006integration} and its calculation is reported in the End Matter \ref{end_RMPS}. 
At large $\chi$, the mutual information simplifies to 
\begin{equation}
    \mathbb{E}\left[I_2(A:B)\right] \approx 
    \log
    \left[ 
    1 + e^{- r \log d } 
    \left( \frac{\chi d}{d+1} \right)^2
    \right]
    \, .
\end{equation}
This expression is clearly finite on a length scale $r \sim \log(\chi^2)$ and it decays to zero for larger values of $r$.
We are then in position to define an \textit{effective correlation length} $\xi_{\rm eff} \sim \log(\chi^2)$, stemming from the sum of overlaps in Eq.~\eqref{eq_mutual_info_TI}, which is the true relevant correlation length at large $\chi$, rather than the correlation length set by the leading eigenvalue $\xi = 1/\log d$. 
That explains why we can find a finite correlation length $\xi$ for an RMPS with arbitrarily large bipartite entanglement $S \sim \log(\chi)$.
The effective correlation length $\xi_{\rm eff}$ diverges with $\chi$, therefore, correlations can still spread arbitrarily far.
Generally, in all MPS ensembles flowing to volume-law states at large $\chi$, we can expect a $\chi$-diverging correlation length. 

 As a side note, we should expect a similar phenomenology in generic random quantum channels. There, by analogy to the correlation length in MPSs, we expect an effective thermalization time to scale as $\log d^N \sim N$, a phenomenon similar to the phantom eigenstate in unitary circuits, see \cite{Bensa2021}, but which seems to have escaped previous studies.

 \emph{Ensembles of MPSs: Unitary and Monitored Dynamics ---}
We introduce different ensembles of MPSs. Consider a quantum circuit evolution where, after each gate application, the time evolution is projected back onto MPSs with a fixed bond dimension $\chi$ by truncating the number of Schmidt values and normalising accordingly. We consider a random quantum system represented by a random unitary circuit with gates drawn from the Haar ensemble, either with or without translation invariance (TI). In the absence of projection onto MPSs, these models are known to exhibit fully quantum chaotic behaviour and lead to volume-law entangled quantum states at large times. Starting from a simple product state, we analyze the spectral density of MPSs, as shown in Fig.~\ref{fig_RMPS_TI_Haar_eigs} and Fig.~\ref{fig_radial_prob_dist}.  

The spectra of the TI-Haar and RMPS ensembles have similar but distinct limiting shapes, both quickly converging to a disk of radius \(1/\sqrt{d}\). By studying the mutual information between two large subsystems, as shown in Fig.~\ref{fig_mutual_info}, which defines the effective correlation length \(\xi_{\rm eff}\), we demonstrate that the two ensembles exhibit different correlation spreading behaviours. Specifically, the effective correlation length diverges as \(\xi_{\rm eff} \sim \log \chi^\alpha\), with \(\alpha_{\rm Haar-TI} \simeq 1.8\), differing from the RMPS value of \(\alpha_{\rm RMPS} = 2\). Notably, this difference is not related to the spectral gap, which is identical for both Haar circuits and RMPSs, but rather to the spectral density, as all eigenvalues contribute to it, see eq.~\eqref{eq_mutual_info_TI}. In particular, Haar circuits (with or without translation invariance) exhibit a higher concentration of small eigenvalues, leading to slower correlation spreading.

\begin{figure}[t!]
  \centering
  \includegraphics[trim=10 10 10 10, clip, width=\columnwidth]{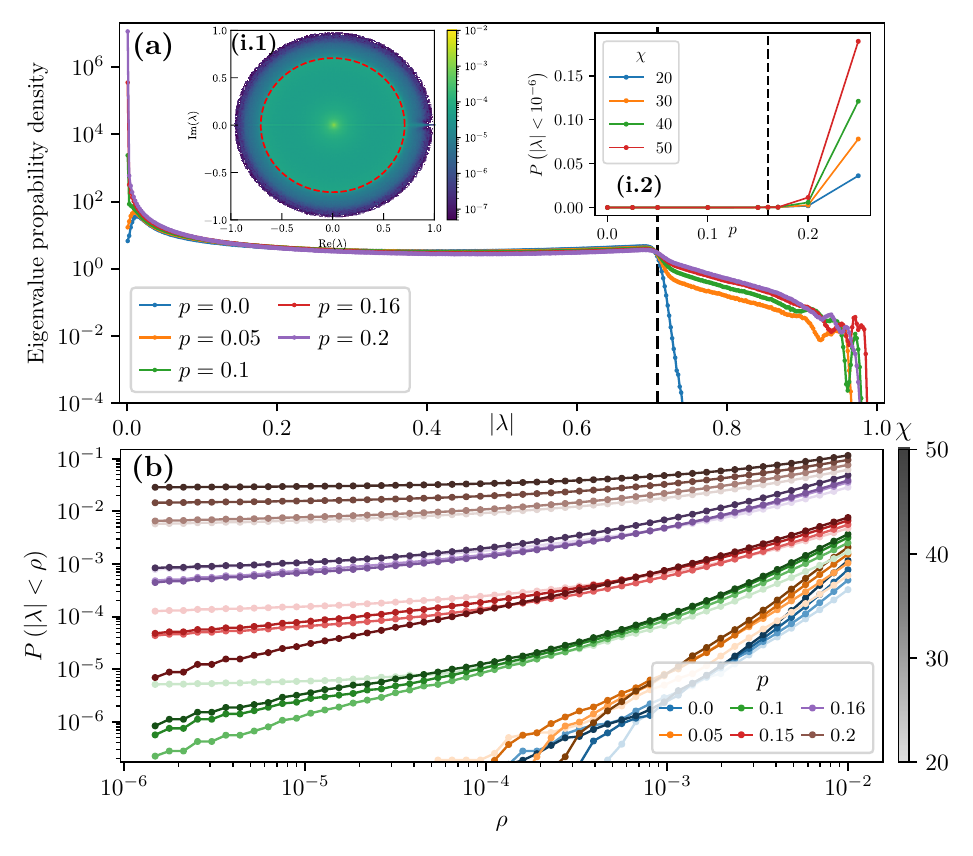}

  \caption{
   (a) Radial probability density of the eigenvalues of the transfer matrix at fixed bond dimension $\chi = 50$ obtained by a monitored circuit of size $N=60$ (eigenvalues at $|\lambda| = 1$ have been removed).
   The dashed line indicates a circle of radius $1/\sqrt{2}$. 
   (i.1) The eigenvalue density in the complex plane for $\chi = 50$ and $p = 0.05$.
   (b) Plot of the probability $P(|\lambda|<\rho)$ of finding an eigenvalue with norm up to  $\rho$ as a function of $\rho$ for different $p$ and $\chi$. 
   For $p \geqslant 0.16$ the curves have been artificially shifted by $\times 5$ for visualization purposes.
   The plot indicates that the limit $\lim_{\rho \to 0^+} P(|\lambda|<\rho)$ is:  for  $p=0.1$, non-zero only for $\chi = 20$ and zero otherwise; for  $p=0.15$ non-zero for $\chi < 50$ and zero otherwise;  for $p \geq 0.16$ non-zero for all $\chi$.  
   (i.2) The same probability computed only for the smallest $\rho$ we considered, as a function of $p$.
  }
  \label{fig_BW_eigs}
\end{figure}

We now turn to the ensemble generated by a random Haar circuit under external monitoring, with a measurement probability \(p\) at each site and layer. This circuit is known to exhibit a MIPT from volume- to area-law scaling at \(p_c \simeq 0.16\)~\cite{zabalo2020criticalpropertiesof,PhysRevX.9.031009,nahum2921measurementandentanglement,jian2020measurementinducedcriticality,10.21468/SciPostPhys.7.2.024,potter2022entanglementdynamicsin,chan2019unitaryprojective,li2019measurementdrivenentanglement,skinner2019measurementinducedphase,czischek2021simulating,han2022entanglementstructure,minoguchi2022continuousgaussianmeasurements,altland2022dynamicsofmeasured,fuji2020measurementinducedquantum,jian2021yangleeedge,bentsen2021measurementinducedpurification,yang2022entanglementphasetransitions,medina2021entanglementtransitionsfrom,lunt2020measurementinducedentanglement,szyniszewski2019entanglementtransitionfrom,tang2020measurementinducedphase,iadecola2022dynamicalentanglementtransition,odea2022entanglementandabsorbing,ravindranath2022entanglementsteeringin,PhysRevLett.130.120402,sierant2023entanglement,buchhold2022revealing,vijay2020measurementdrivenphase,fan2021selforganizederror,li2021conformal,ippoliti2021entanglementphasetransitions,turkeshi2020measurementinducedcriticality,turkeshi2022measurementinducedcriticality,sierant2022measurementinducedphase,zabalo2022infiniterandomnesscriticality,2023Google,altman2023}. 
In the volume-law phase (\(p < p_c\)), the correlation length grows as \(\log \chi^\alpha\), which we confirm in Fig.~\ref{fig_mutual_info}. 
The effective correlation length diverges with an exponent \(\alpha(p)\) that depends on the measurement rate \(p\) and decreases as \(p\) approaches the critical measurement rate \(p_c\).
Moreover, we observe that the spectral density is significantly altered compared to the unitary case: the presence of measurements introduces additional (circular) tails outside the circle of radius \(1/\sqrt{d}\), even for small measurement rates, see Figs.~\ref{fig_BW_eigs}(a) and \ref{fig_BW_eigs}(i.1).
However, this does not strongly affect physical correlations: in such circular spectral densities, the contributions of states with small gaps effectively cancel out in the sum   \eqref{eq_mutual_info_TI} as dephasing angles.   The most relevant feature of the spectral density near the MIPT is not the density of eigenvalues close to \(1\), but rather the one of eigenvalues close to zero. A large population of vanishing eigenvalues leads to slower correlation spreading, as fewer non-zero terms contribute to the sum in Eq.~\eqref{eq_mutual_info_TI}.
Focusing on the probability of eigenvalues with absolute values smaller than \(\rho\), $P(|\lambda|<\rho)$, shown in Figs.~\ref{fig_BW_eigs}(b) and \ref{fig_BW_eigs}(i.2), we observe that, while in the volume-law phase (\(p < p_c(\chi)\)), this concentration decays to zero as \(\rho \to 0^+\), in the area-law phase (\(p > p_c(\chi)\)) converges instead to a finite value. In this latter case, similarly to the MPS representation of the ground state of gapped Hamiltonian, we indeed expect that by increasing $\chi$, an increasing number of Schmidt values and transfer matrix eigenvalues are vanishing small, as the description of the state is exact (up to exponentially small deviations) for  $\chi> \bar{\chi}(p)$, and correlations stop spreading, namely $\xi_{\rm eff} \sim \log  \chi^{\alpha=0}$, which we indeed confirm in Fig.~\ref{fig_BW_alpha}.
Using the value $P(|\lambda|<0^+)$ as the effective order parameter, by extrapolating the limit $\rho \to 0 ^+$ in Fig. \ref{fig_BW_eigs}(b),  we find, quite remarkably, that  \textit{the transition is present for any value of $\chi$}, with a critical measurement rate $p_c(\chi)$ converging from the left as  $p_c(\chi) \rightarrow_{\chi \to \infty} p_c \simeq 0.16$. We have therefore shown that the MIPT can be detected also in circuits where the time evolution of the state is projected on an MPS with finite bond dimension.

\emph{Conclusions ---}  
We studied ensembles of matrix product states (MPSs) from the perspective of the spectrum of the transfer matrix and its associated correlation length. Our findings reveal that, for states generated by strongly entangling quantum dynamics, the standard correlation length defined in terms of the spectral gap is not physically relevant. Instead, a new effective correlation length emerges, governed by the combined contribution of multiple eigenstates of the transfer matrix. 

In generic (chaotic) unitary quantum circuits generating MPSs, the spectrum is confined within a circle of radius \(1/\sqrt{d}\), as in random quantum channels. However, the spectral density, and particularly the concentration of small eigenvalues, is highly model-dependent. This concentration determines an effective correlation length, \(\xi_{\rm eff}\), which grows as \(\log \chi^\alpha\), where \(\alpha\) is a model-dependent exponent that encodes physical information about the underlying dynamics. Importantly, \(\xi_{\rm eff}\) determines the effective scales over which correlations remain significantly non-zero.

Our results provide two key insights. First, MPSs generated by truncating dynamical quantum evolution require a fundamentally different paradigm compared to those near critical ground states, as their spectral densities are radially uniform, and not peaked around the frequencies of the quasiparticle spectra \cite{zauner2015transfer}. Second, despite the reduced complexity of MPSs, having bond dimension much smaller than exponential with system size, they still retain meaningful physical properties of the underlying quantum dynamics.
This includes the ability to capture the spread of the entanglement and detect entanglement transitions at any finite value of $\chi$. While, on one hand, this opens new ways to study MIPTs and generically entanglement dynamics under external monitoring or dissipation, possibly in simple tensor models where analytical treatments are possible,  our findings also open new research directions in the study of dynamical phases using tensor networks. They suggest the possibility of extending equilibrium tensor network techniques~\cite{PhysRevLett.123.250604,PhysRevLett.129.200601,PhysRevB.78.024410,2411.03954,PhysRevLett.102.255701} to the study of non-equilibrium phases of matter, potentially even in higher dimensions using PEPS tensor networks.

\begin{acknowledgments}
\paragraph{Acknowledgments. ---}  
We wish to thank L. Tagliacozzo, C. Lancien, M. McGinley and A. De Luca for inspiring discussions and for collaborations on topics connected with this work. This work was founded by the ERC Starting Grant 101042293 (HEPIQ) and the ANR-22-CPJ1-0021-01.  This work was granted access
to the HPC resources of IDRIS under the allocation
AD010914149R1 and with MPS codes developed also using the C++ iTensor library \cite{itensor-r0.3}. 
\end{acknowledgments}

\bibliography{references.bib}
\bibliographystyle{apsrev4-2}
\clearpage
\setcounter{section}{0}
\setcounter{secnumdepth}{2}

\appendix
\renewcommand\thefigure{\thesection.\arabic{figure}}    

\begin{center}
    \textbf{End Matter}
\end{center}


\setcounter{figure}{0}

\section{Calculation of the mutual information in RMPS}
\label{end_RMPS}

RMPSs can be written as MPSs where each site is associated with a Haar-random unitary matrix of size $d \chi \times d \chi$ (in the case where all matrices are identical, this is referred to as an iRMPS):
\begin{equation}\label{eq_RMPS_tensor}
\begin{gathered}
    \includegraphics[width=0.4\columnwidth]{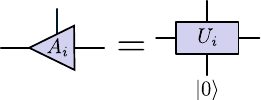}
\end{gathered}
\,.
\end{equation}
Vertical links have dimension $d$, while horizontal ones have dimension $\chi$.
The triangle shape of the RMPS tensor $A_i$ signals that the identity matrix is a right eigenvector of the transfer matrix $\mathcal{T}_i$ with eigenvalue 1, which stems from the unitarity of $U_i$, meaning that the state is properly normalized by construction. 

The averaged replicated transfer matrix can be computed as follows 
\begin{equation}\label{eq_rmps_tm_bond}
\begin{aligned}
    \mathcal{T}_\alpha^{(k)} & = 
    \mathbb{E} \left\{
    \begin{gathered}
        \includegraphics[width=0.2\columnwidth]{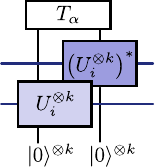}
    \end{gathered}
    \right\} 
    =
    \begin{gathered}
        \includegraphics[width=0.4\columnwidth]{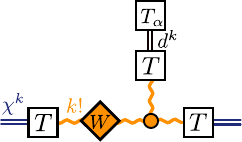}
    \end{gathered}
\end{aligned}
    \,,
\end{equation}
where $W$ refers to the Weingarten matrix $W_{\sigma \pi} = \Wg(\sigma^{-1} \pi, d)$, the orange dot is the copy tensor, and tensors $T$ are representations of permutations acting on $k$ indices (grouped in the diagram). 
The dimensions of the links are explicitly written, where the blue lines correspond to replicated MPS bond indices, the black lines correspond to replicated physical indices, and the wavy orange lines correspond to permutation indices.
The size of the transfer matrix in Eq.~\eqref{eq_rmps_tm_bond} grows polynomially with $\chi$ but that can be avoided if we instead work with the shifted transfer matrices
\begin{equation}\label{eq_tm_rmps_permutation}
    \mathcal{\Tilde{T}}_\alpha^{(k)} = 
    \
    \begin{gathered}
        \includegraphics[width=0.4\columnwidth]{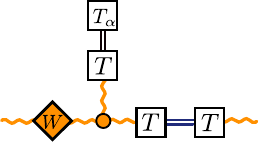}
    \end{gathered}
    \,,
\end{equation}
of dimension $k!$.

The permutation $\alpha$ is fixed as explained in the main text, so we define the transfer matrices $\mathcal{T}_{(1)} = \mathcal{\Tilde{T}}_e^{(k)}$ and $\mathcal{T}_{(2)} = \mathcal{\Tilde{T}}_{C_k}^{(k)}$.
For $k=2$, the transfer matrices are $2\times2$ matrices which can be readily computed and diagonalized (see also \cite{PRXQuantum.4.030330}).

\begin{figure}[t]
    \centering
    \includegraphics[width=\columnwidth]{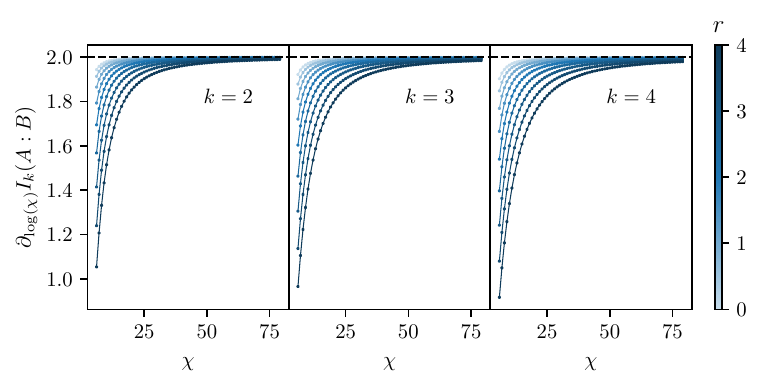}
    \caption{
    Partial derivate with respect to $\log(\chi)$ of the Rényi-$k$ mutual information in an RMPS, for $k \in \{2,3,4\}$.
    The horizontal dashed line corresponds to the analytically computed value at large $\chi$ for $k = 2$. In all cases, we confirm the exponent $2$ for the spreading of the mutual information. 
    }
    \label{fig_rmps_Ik_grad}
\end{figure}

The standard Von Neumann mutual information can be recovered from the analytical continuation of Rényi to $k = 1$.
Computing $I_k$ for $k>2$ generically becomes prohibitively numerically expensive very rapidly with increasing $\chi$. 
However, in the RMPS case, the transfer matrices given by Eq.~\eqref{eq_tm_rmps_permutation} can be easily numerically diagonalized for any $\chi$ and not too large $k > 2$, and $I_k$ can be computed from Eq.~\eqref{eq_mutual_info_TI}.
If we assume that the mutual information spreads as $I_k \sim \log(\chi^\alpha)$,
$\alpha$ can be found by computing the partial derivative $\partial_{\log \chi}
I_k$, as is shown in In Fig.~\ref{fig_rmps_Ik_grad}.
We conclude that, for any $k \in \{2,3,4\}$, the exponent $\alpha$ tends to $2$ as $\chi$ is increased.
Therefore, the analytical continuation to $k= 1$ is trivial, i.e. $I \sim \log(\chi^2)$, so we expect the same effective correlation length for the Von Neumann mutual information.

\section{Correlation spreading in the monitored Haar circuit}
\setcounter{figure}{0}    

\begin{figure}
    \centering
    \includegraphics[width=0.8\columnwidth]{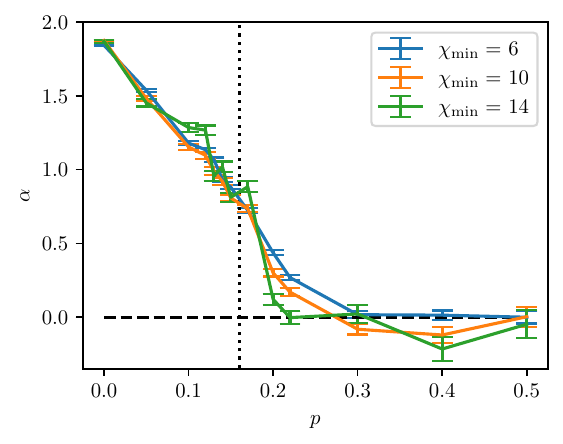}
    \caption{
    Spreading exponent of the effective correlation length in the BW monitored circuit, as a function of measurement probability $p$.
    The exponents are found by fitting the mutual information as a function of $\chi$ with minimal bond dimension $\chi_{\textnormal{min}}$.
    The vertical dashed line signals the critical $p_c=0.16$ for the MIPT.
    }
    \label{fig_BW_alpha}
\end{figure}

As seen in Fig.~\ref{fig_mutual_info}, at large $\chi$, the mutual information spreads with effective correlation length $\xi_{\textnormal{eff}} = \log(\chi^\alpha)$, where the exponent $\alpha$ can be found with a linear fit.
In Fig.~\ref{fig_BW_alpha} we show how the spreading exponent depends on the measurement probability $p$.
The results reflect the presence of the MIPT, since for $p > p_c$ the spreading exponent converges to 0, as the minimal fitting bond dimension $\chi_{\textnormal{min}}$ is increased.
The exponent $\alpha = 0$ agrees with the area-law phase since correlations have finite length.
Instead for $p < p_c$, the exponents converge to a finite value, reflecting the fact that in the volume-law phase, the effective correlation length is expected to diverge.

\section{Asymptotic spectral comparisons}
\setcounter{figure}{0}    

\begin{figure}
    \centering
    \includegraphics[width=0.9\columnwidth]{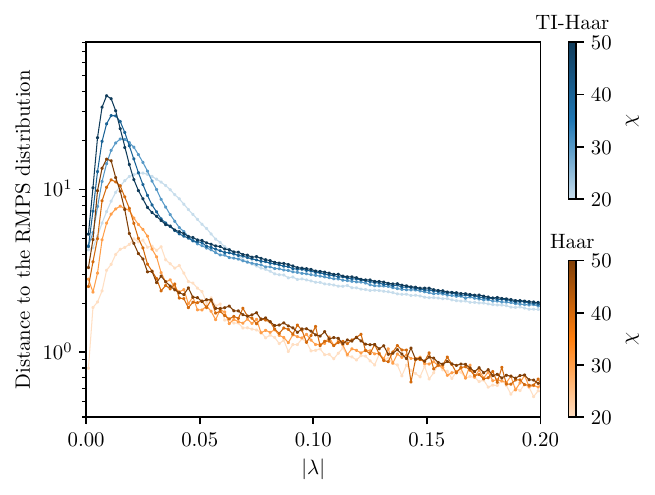}
    \caption{Difference between the radial spectral density of the RMPS and the ones of the TI-Haar circuit (blue) and the Haar circuit (orange), as a function of eigenvalue norm $|\lambda|$, for different fixed bond dimensions $\chi$. We observe a convergence to a positive-valued function, signalling in both cases the presence of a larger number of small eigenvalues compared to the RMPS case. }
    \label{fig_radial_prob_dist}
\end{figure}

In Fig.~\ref{fig_RMPS_TI_Haar_eigs} we compare the radial spectral density of the RMPS and the TI-Haar circuit. 
Although they appear asymptotically different, in Fig.~\ref{fig_radial_prob_dist} we do a more rigorous analysis by plotting the difference between the spectrums, also including the regular (non-TI) Haar circuit. 
We conclude that the distribution of eigenvalues is more similar to the RMPS in the non-TI case than in the TI case of the Haar circuit.
We postulate that this difference occurs because the spectrum of the transfer matrix is sensitive to the system's translational invariance.
However in both cases, asymptotically, the spectral densities do not converge to the RMPS one, indicating that the spectrum is sensitive to the microscopic details of the system even for ergodic, highly truncated dynamics.

%
%

\end{document}